\documentclass{article}
\usepackage{arxiv}

\usepackage[utf8]{inputenc} % allow utf-8 input
\usepackage[T1]{fontenc}    % use 8-bit T1 fonts
\usepackage{hyperref}       % hyperlinks
\usepackage{url}            % simple URL typesetting
\usepackage{booktabs}       % professional-quality tables
\usepackage{amsfonts}       % blackboard math symbols
\usepackage{nicefrac}       % compact symbols for 1/2, etc.
\usepackage{microtype}      % microtypography
\usepackage{cleveref}       % smart cross-referencing
\usepackage{lipsum}         % Can be removed after putting your text content
\usepackage{graphicx}
\usepackage[numbers]{natbib}
\usepackage{doi}
\usepackage{lscape} % 途中から部分的に用紙を横に使いたいしたいときはlscapeパッケージを入れて(\usepackage{lscape})、横にしたい所をlandscape環境でくくる

\title{
	Odor Map by LLM Inference of Verbal Descriptions 
}

\newif\ifuniqueAffiliation

\uniqueAffiliationtrue
\ifuniqueAffiliation % Standard variant of author block
\author{ \\
	\href{https://orcid.org/0009-0000-4254-7803}{\includegraphics[scale=0.06]{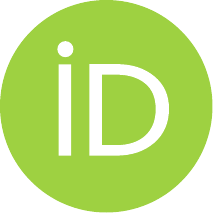}\hspace{1mm}Yuki Harada} \\% Name + ORCID
		Institute of Integrated Research, Institute of Science Tokyo, \\
		4259, Nagatsuta-cho, Midori-ku, Yokohama, Kanagawa 226-8501, JP \\
	\texttt{harada.y.ai@m.titech.ac.jp} \\
	\And
	\href{https://orcid.org/0000-0003-0620-2757}{\includegraphics[scale=0.06]{fig/orcid.pdf}\hspace{1mm}Manuel Aleixandre} \\% Name + ORCID
		Institute of Integrated Research, Institute of Science Tokyo, \\
		4259, Nagatsuta-cho, Midori-ku, Yokohama, Kanagawa 226-8501, JP \\
	\texttt{aleixandre.m.aa@m.titech.ac.jp} \\
	\And
	\href{}{\hspace{1mm}Manabu Okumura} \\% Name
		Institute of Integrated Research, Institute of Science Tokyo, \\
		4259, Nagatsuta-cho, Midori-ku, Yokohama, Kanagawa 226-8501, JP \\
	\texttt{oku@pi.titech.ac.jp} \\
	\And
	\href{https://orcid.org/0000-0002-0599-226X}{\includegraphics[scale=0.06]{fig/orcid.pdf}\hspace{1mm}Takamichi Nakamoto} \\% Name + ORCID
		Institute of Integrated Research, Institute of Science Tokyo, \\
		4259, Nagatsuta-cho, Midori-ku, Yokohama, Kanagawa 226-8501, JP \\
	\texttt{nakamoto.t.ab@m.titech.ac.jp} \\
}

\else
\usepackage{authblk}

\newbox{\orcid}\sbox{\orcid}{\includegraphics[scale=0.06]{fig/orcid.pdf}} 
\author[1]{%
	\href{https://orcid.org/0009-0000-4254-7803}{\usebox{\orcid}\hspace{1mm}Yuki Harada\thanks{\texttt{harada.y.ai@m.titech.ac.jp}}}
}
\author[1]{%
	\href{https://orcid.org/0000-0003-0620-2757}{\usebox{\orcid}\hspace{1mm}Manuel Aleixandre\thanks{\texttt{aleixandre.m.aa@m.titech.ac.jp}}}
}
\author[1]{%
	\href{https://orcid.org/}{\usebox{\orcid}\hspace{1mm}Manabu Okumura\thanks{\texttt{oku@pi.titech.ac.jp}}}
}
\author[1]{%
	\href{https://orcid.org/0000-0002-0599-226X}{\usebox{\orcid}\hspace{1mm}Takamichi Nakamoto\thanks{\texttt{nakamoto.t.ab@m.titech.ac.jp}}}
}
\affil[1]{ \\
	Institute of Integrated Research, Institute of Science Tokyo, \\
	4259, Nagatsuta-cho, Midori-ku, Yokohama, Kanagawa 226-8501, JP \\
}
\fi

\renewcommand{\shorttitle}{
	Odor Map by LLM Inference of Verbal Descriptions 
}

\hypersetup{
pdftitle={
	Odor Map by LLM Inference of Verbal Descriptions 
},
pdfsubject={
	汎用LLMを利用した官能評価の推論生成と匂いマップの作製, 原田祐希、Manuel Aleixandre、奥村学、中本高道
},
pdfauthor={
	Yuki Harada, 
	Manuel Aleixandre, 
	Manabu Okumura, 
	Takamichi Nakamoto
},
pdfkeywords={Odor space, Odor map, large language model (LLM), digital twin, automatic scent creation, Dravnieks},
}

\begin{document}
\maketitle

\begin{abstract}

Odor maps that accurately reflect both the diversity and specificity of human olfaction are essential for advancing digital scent technology. 
The digitalization of odors and virtual reality (VR) implementation in olfaction have lagged due to the difficulty of archiving them. 
If such data were accessible, it would accelerate the development of digital scent technologies. 
In this study, 
we present a method that provides both verbal descriptions of odors and corresponding odor maps. 
We here performed statistical tests to compare the large language model (LLM) inferences of verbal descriptions with the existing LLM-derived similarity scores 
and with sensory metrics collected in 1985 by Dravnieks. 
These tests revealed that the consistency of the inferences of verbal descriptions was comparable to that of the existing LLM-derived similarity scores. 
Furthermore, we found that prompt optimization could reduce inconsistencies in verbal descriptions. 
This technology here can contribute to olfactory VR since the intended scent in the olfactory content can be easily generated based on odor map. 
% CurrentSubmission.260617_1530.pdf

% \section*{Author summary}
% 
% Authers' group has investigated automatic scent creation. 
% The key objective is to translate odor space into a sensory evaluational one. 
% This study demonstrates a potential to generate odor maps by LLM-derived pairwise similarities of odor names. 
% We believe LLMs hold significant promise for advancing digital scent technology and overcoming the limitations of traditional, large-scale sensory evaluation methods.

\end{abstract}

% keywords can be removed

% teaserfigure (Fig. 1) Odor Map based on verbal descriptions inferred by GPT-4o-mini. 
\begin{figure*} % {figure}[!t]% 全幅画像挿入
  \includegraphics[width=\textwidth]{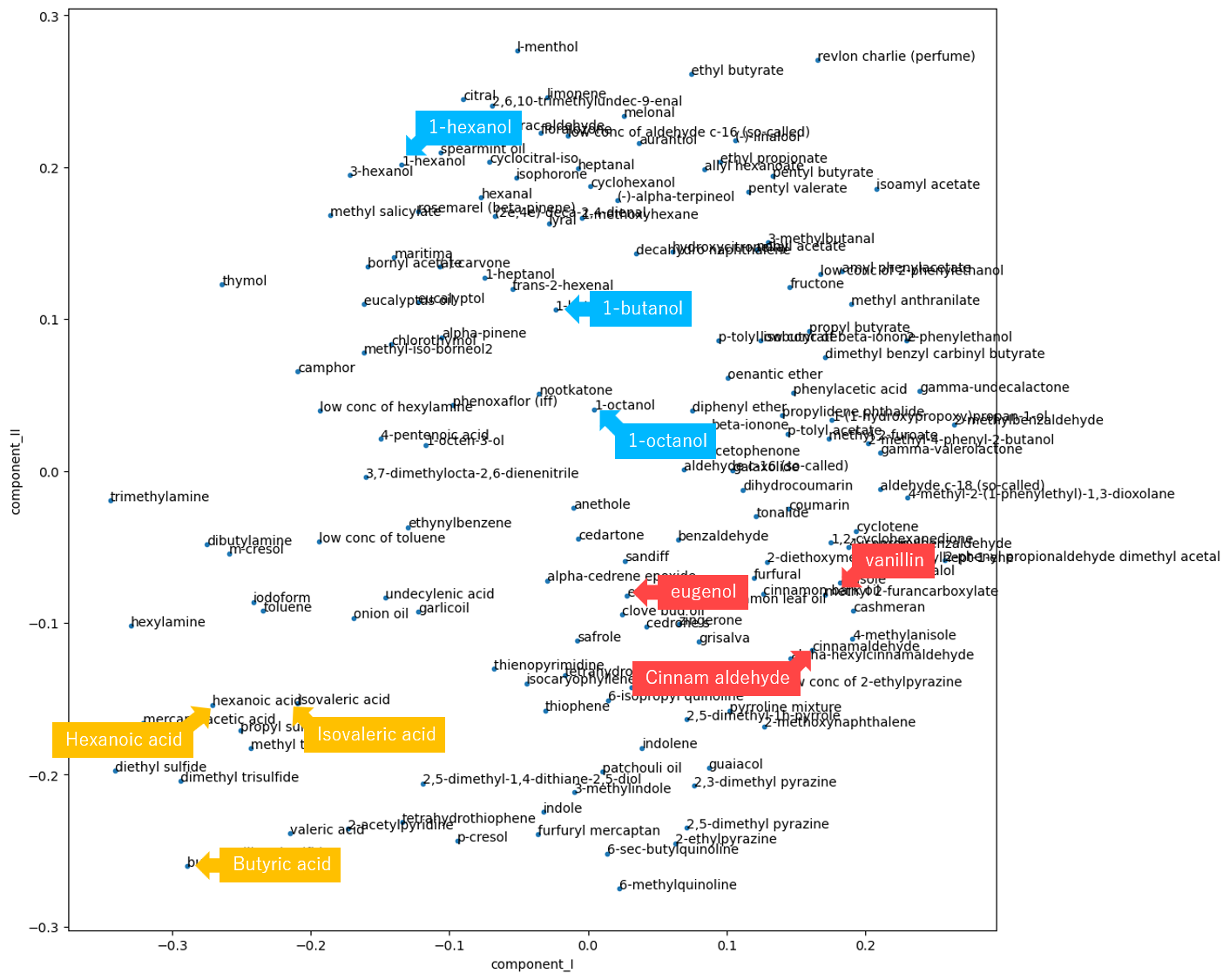}
  \caption{Odor Map based on verbal descriptions inferred by GPT-4o-mini. 
  This 2D representation of odors was created using MDS (Multi-dimensional Scaling) based on pairwise distances between embeddings of each name's LLM inference.
  Similar odors are expected to have coordinates that are close together.
  In this map, alcohols are highlighted in light blue, carboxylic acids in yellow, and aromatics in red.
  We found that chemicals with the same functional groups tend to be located closely together. }
%   \Description{}
\end{figure*} % {figure}

%% ======================== Main % ======================== 
\maketitle

\section{Introduction}

Compared to other senses, digitalization of odors and VR implementation for olfaction lags behind. 
To describe the quality of an odor/scent in language, human sensory evaluation is necessary. 
The sensory evaluations require evaluators to undergo extensive training to develop their ability to distinguish between smells and express them verbally. 
Therefore, obtaining large-scale, high-quality sensory evaluations remains a significant challenge.

Digital scent technologies is in early stage but is now progressing rapidly \cite{nakamoto2025digital}. 
Several researchers study smell-perception prediction from molecular structural parameters and mass spectrum 
\cite{khan2007predicting,keller2017predicting,gutierrez2018predicting,ravia2020measure,lee2023principal,satarifard2025high,nozaki2016odor}. 
Moreover, algorithmic scent creation using AI technologies was studied \cite{aleixandre2024automatic,aleixandre2025generative,aleixandre2026system}. 
Although large-scale sensory data are indispensable for the development, 
they need huge amout of time and labor. 
The sensory dataset made by Dravnieks has been frequently used in this field \cite{dravnieks1985atlas}. 
It was made about 40 years ago and the scale of the dataset is not enough for recent study of digital scent technologies. 
Large-scale sensory data of olfaction can rapidly accelerate digital scent technologies if we obtain it easily. 
% digital olfaction techniques

Kurfali showed the ability of Large Language Models (LLMs) to recover sensory information from natural language \cite{kurfali2025representations}.
Then, they reported the possibility of LLM-derived similarity scores using short prompts. 
They compared it with Dravnieks' dataset and then found 
% that statistically comparing these similarity scores with pairwise distances from the original sensory evaluation data revealed 
that LLMs can infer odor similarity to some degree  
after comparing these similarity scores with pairwise distances from the original sensory evaluation data
\cite{kurfali2025representations}.

Based on the substantial pairwise distance correlation between Dravnieks' data 
and LLM-derived ones as well as the relationship between Stress and the dimension obtained from MDS (Multi-Dimensional Scaling) \cite{dillon1984multivariate} method, 
it was reported that desirable odor maps can be generated using LLMs such as GPT-4o-mini and Gemma 3:12b \cite{harada2026odor}. 
It was also reported in the paper that odor map using LLM-derived similarity scores offer a valuable solution for overcoming the limitations of sensory tests.

% However, these methods, based on LLM-derived similarity scores, have two disadvantages: 
% first, detailed verbal sensory descriptions are not realized; % FIX
% second, the resulting odor map coordinates lack invariance. 
% When compared to static models like Word2Vec and FastText reported in 2018 \cite{nozaki2018correction}, 
% which required a corpus trained specifically on olfactory data but exhibited relatively low invariance (robustness), 
% current similarity-based approaches can only represent relative coordinates within a specific odor space representation. 
% This study explores the potential of using LLMs 
% to infer verbal sensory descriptions and obtain their embedding representations, potentially replacing large-scale sensory evaluations.

However, that method based on LLM-derived similarity scores has room for improvement. 
First we do not obtain detailed sensory descriptions. 
We do not know how LLM interprets each odor even if we obtain similarities between odors. 
It is better to have detailed sensory description from the view point of explainability. 

Second, the resulting odor map lacks invariance of coordinates. 
The coordinate of each odor depends on a set of odors used in the analysis. 
Since the obtained coordinate is just relative, there are many other candidates obtained by shift and rotation in high-dimensional odor space. 
This situation is similar to static models like Word2Vec and FastText \cite{nozaki2018correction}. 
The similarity-based approach can only represent relative coordinate within limited odor space. 
This study explores more universal method by obtaining the embedding representation of each odor. 
It is more generaalized knowledge since each odor coordinate is not influenced by others. 

This technology here can contribute to olfactory VR \cite{sunami2025exploring,ranasinghe2019tainted} 
since the intended scent in the olfactory content can be easily generated based on odor map.

\section{Materials and methods}

We comparatively verified four representations of odor spaces, including three representations inferred by an LLM. 

% \begin{itemize}
%   \item DravOrig; Sensory evaluational scores in Dravnieks dataset 
%   \item SimByLLM; LLM-derived similarity scores 
%   \item GenSE; Inference of odor verbal descriptions (with unoptimized prompt) 
%   \item OptGenSE; Inference of odor verbal descriptions (with DSPy-optimized prompt) 
% \end{itemize}

\subsection{DravOrig; Sensory evaluational scores in Dravnieks dataset } %  1. 

% First representation of odor space is descriptor's suitability scale: 
In 1985, Andrew Dravnieks published a semantic profiling of 180 odorants \cite{dravnieks1985atlas}. 
His data used here were taken from pyrfume dataset \cite{castro2022pyrfume, githubpyrfume}. 
A list of 146 odor descriptors 
selected from an initial list of 800, was provided by over 100 panel members from different organizations. 
These members then evaluated each descriptor's suitability on a scale of 0 to 5 for 160 odorant.

% Fig. 2  Odor Map based on Dravnieks' sensory metrics (DravOrig). 
% \begin{landscape}
\begin{figure*} % {figure}[!t]% 全幅画像挿入
    \centering
    \includegraphics[keepaspectratio, width=5.2 in ]{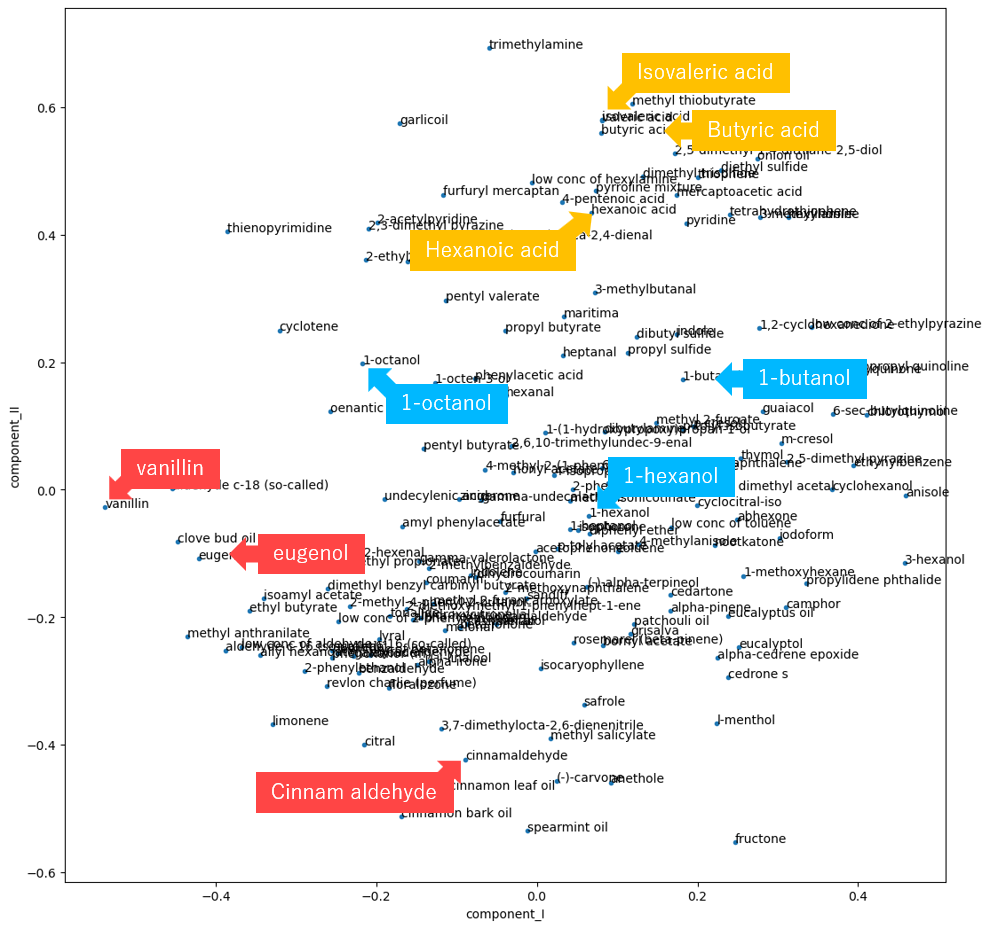}
    \caption{Odor Map based on Dravnieks' sensory metrics (DravOrig). 
  This 2D mapping was created using MDS (Multi-dimensional Scaling) based on pairwise cosine distances between embeddings of each sensory evaluation in Dravnieks.
  In this map, alcohols are highlighted in light blue, carboxylic acids in yellow, and aromatics in red.}
\end{figure*} % {figure}
% \end{landscape}

% Odor Map based on Dravnieks' sensory metrics (DravOrig). 
% This 2D mapping was created using MDS (Multi-dimensional Scaling) based on pairwise cosine distances between embeddings of each sensory evaluation in Dravnieks.
% In this map, alcohols are highlighted in light blue, carboxylic acids in yellow, and aromatics in red.
% "C:\Users\yharada\Documents\CurrentTex.260601.DSPyGenSE.acm_vrst2026\fig\MDS2D_dravorig.png"

\subsection{SimByLLM; LLM-derived similarity scores}

A Second representation of odor space is LLM's prediction of the similarity. 
Pairwise similarity scores were inferred by an LLM, following a prompting approach established in the previous work \cite{kurfali2025representations,harada2026odor}. 
We generated these pairwise similarity scores for 160 odor names drawn from the Dravnieks dataset, 
resulting in a total of 12,720 individual similarity scores. 
This extensive set of scores was then used as input to our Multidimensional Scaling (MDS)-derived odor map. 
The MDS analysis visualizes and spatially arranges odors based on these predicted similarities. 
Prompt is shown in appendix A1.

\subsection{GenSE and OptGenSE; Inference of odor verbal descriptions } %3

Verbal descriptions, reflecting the sensory evaluations, provide another representation of odor space. 
We consider that LLM inferences of odor verbal descriptions represent a potential model of odor perception. 
We performed two types of verbal description inference. 
The first inference (GenSE) utilized a few-shot prompting approach associated with each odor, 
generating verbal descriptions for all 160 odorants listed in the Dravnieks dataset. 

The second inference (OptGenSE) utilized a two-stage agent framework to generate comprehensive sensory evaluations. 
Here, an odor name serves as input to create a detailed odor description, 
which is subsequently used to generate an overall sensory evaluation. 
Each agent within this framework was optimized using DSPy 
and utilized to obtain verbal descriptions for the 160 odorants listed in the Dravnieks dataset. 
We utilized the Python DSPy package \cite{khattab2023dspy} to optimize the prompting sequence, and to execute the resulting prompts.

The prompts for GenSE and OptGenSE are shown in A.2 and A.3, respectively.

% \begin{landscape}
\begin{table*} % {figure}[!t]% 全幅画像挿入
    \centering
    \caption{A sample of verbal output in GenSE and OptGenSE, 
    demonstrates that prompt optimization using DSPy could produce more reasonable verbal descriptions. }
    \includegraphics[keepaspectratio, width=\textwidth ]{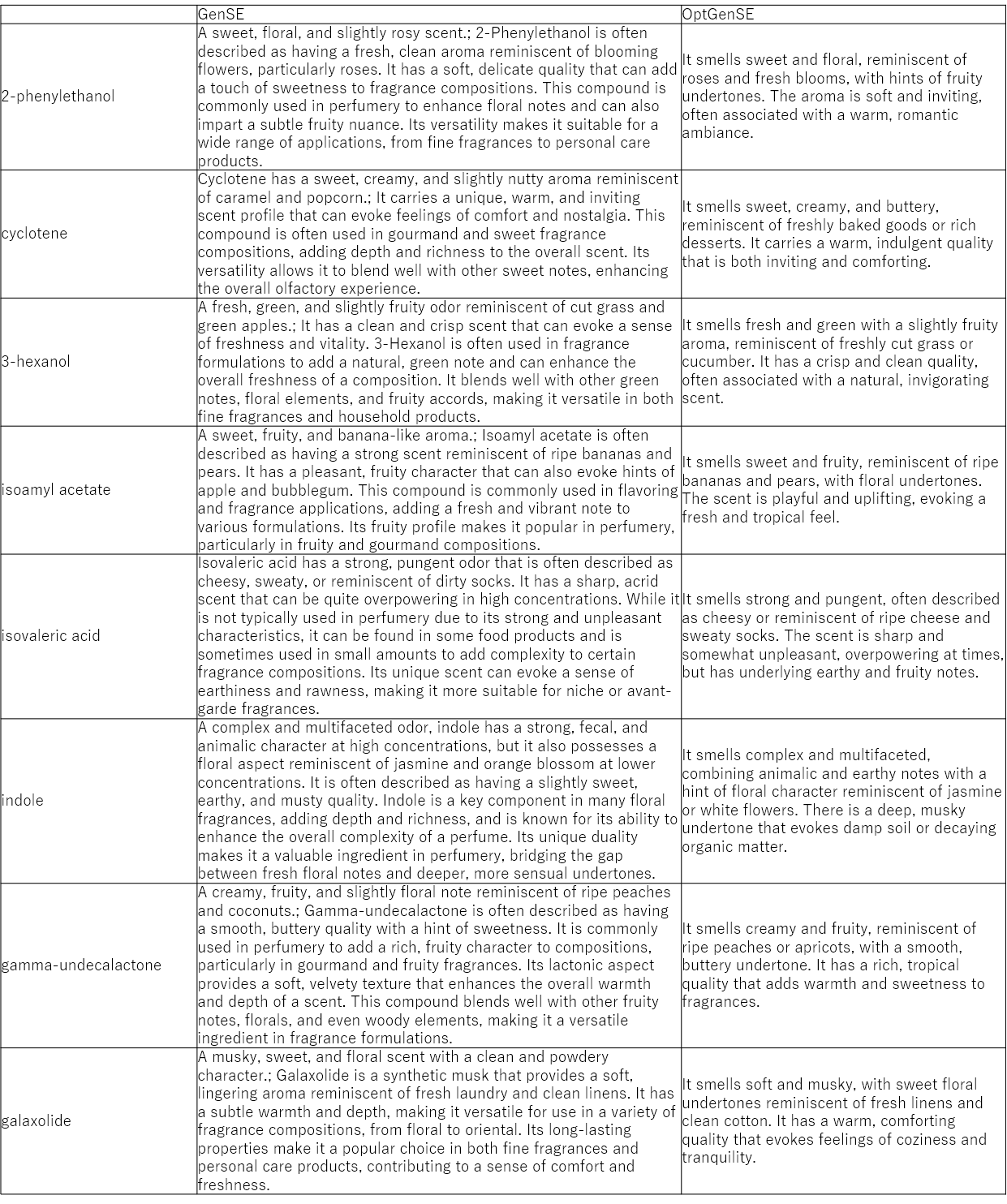} % \textwidth
\end{table*} % {figure}
% \end{landscape}

\subsection{Distance Measures and Statistical Analysis}

We conducted a Mantel test \cite{mantel1967detection} to compare distance measures across four different representations of odor space, 
i.e., DravOrig, SimByLLM, GenSE and OptGenSE. 
The Mantel test compares two distance matrices by calculating the correlation between distances 
in their lower (or upper) triangular portions. 
We used the Python library `skbio.stats.distance.mantel` \cite{skbiomantel} to perform this statistical analysis.

To enable statistical analysis of pairwise distances between odor names in these four odor spaces, 
we converted the verbal descriptions in GenSE or OptGenSE into vector representations. 
Following previously reported research \cite{zhong2024sniff} and commonly employed in Retrieval-Augmented Generation (RAG), 
we utilized embedding models to generate these vectors for both GenSE and OptGenSE. 
Specifically, we calculated pairwise cosine distances between embeddings produced by embeddinggemma:300m \cite{embedding_gemma_2025} 
and subsequently performed statistical analysis.

For DravOrig, cosine distances were calculated based on the numerically evaluated suitability of 160 descriptors. 
% We also employed cosine distances to determine pairwise distances for DravOrig, 
% where 160 descriptor's suitability were numerically evaluated.
The resulting 12,720 pairwise cosine distances will be used in subsequent statistical tests and MDS 
to analyze the coordinates across these four representations.
Fig. 2  illustrates the two-dimensional mapping generated using metric MDS based on DravOrig. 
Pairwise cosine distances were calculated for the 160 odor names in the Dravnieks dataset (a total of 12,720 comparisons).
It is also discussed in section 3.4. 

% % Fig. 5 Correlation coefficients by statistical tests (Mantel's test)  
% % \begin{landscape}
% \begin{figure} % {figure*}[!t]% 全幅画像挿入
%     \centering
%     \includegraphics[keepaspectratio, width=3 in ]{fig/MantelCorrrHeat_4_Representation.png }
%     \caption{Correlation coefficients by statistical tests (Mantel's test); 
%     Color indicates the correlation coefficient between pairwise distances across the four metrics (DravOrig, SimByLLM, GenSE and OptGenSE), 
%     with blue tones indicating higher coefficients and red tones signifying lower ones. 
%     Each cell displays a statistical test result (top), p-value (middle), and significance indicator (bottom). }
% \end{figure} % {figure*}
% % \end{landscape}

\section{Results and Discussion}

The three inferred representations (GenSE, OptGenSE and SimByLLM) were generated by GPT-4o-mini via the OpenAI API.

\subsection{Verbal outputs in GenSE and OptGenSE } 

A sample of verbal descriptions in GenSE and OptGenSE, 
generated through inference for odorants from the Dravnieks dataset, is shown in Table 1. 
The descriptions in Table 1 demonstrate that both GenSE and OptGenSE can produce reasonably descriptive accounts of odors. 
For example, it is well known that isoamyl acetate has smell of banana. 
Table 1 shows that LLMs can infer it appropriately. 
Moreover, isovaleric acid is known to have bad smell associated with sweaty socks. 
Inference by LLM agrees with it. 
Moreover, it was found that OptGenSE generates more compact sentences than GenSE, excluding the redundant sentences.

% Table 2 Correlation coefficients by statistical tests (Mantel's test)  
\begin{table}[t]
\caption{Correlation coefficients by statistical tests (Mantel's test); 
    indicating the correlation coefficient between DravOrig and the pairwise distance of SimByLLM, GenSE and OptGenSE
    % , with blue tones indicating higher coefficients and red tones signifying lower ones. 
    % Each cell displays a statistical test result (top), p-value (middle), and significance indicator (bottom)
    .}
\label{tab:embedding_models}
\centering
\begin{tabular}{p{3.5cm} p{1.8cm} p{1.5cm}}
\toprule
Metrics &
Correlation coefficient &
p-value (mantel)	 \\
\midrule

OptGenSE(+EG300m) &
0.345 &
0.001  \\

GenSE(+EG300m) &
0.295 &
0.001  \\

SimByLLM &
0.332 &
0.001  \\

\bottomrule
\end{tabular}
\end{table}

\subsection{Statistical Comparison of Inferred Representations with Dravnieks' Sensory Metrics} 
% Statistical Comparison between DravOrig and the Three Inferred Representations: SimByLLM, GenSE, and OptGenSE 

To compare the three inferred representations (SimByLLM, GenSE, and OptGenSE) 
with the sensory metrics obtained by Dravnieks (DravOrig), 
we used Mantel tests to compare pairwise distances 
calculated using cosine similarity on embeddings from the 'embeddinggemma:300m' model.
The odor metrics were derived as follows:

\begin{itemize}
  \item DravOrig + cos; Sensory evaluational metrics in Dravnieks dataset
  \item SimByLLM; A set of pair similarities inferred by an LLM
  \item GenSE + EG300m + cos; A set of sensory evaluation texts generated by an LLM (without prompt optimization) 
  \item OptGenSE + EG300m + cos; A set of sensory evaluation texts generated by an LLM (with prompt optimization using DSPy) 
\end{itemize}

Here, 'EG300m' refers to the pre-trained embedding model ('embeddinggemma:300m'), 
and 'cos' denotes the cosine distances used for pairwise distance calculation. 

% We obtained correlation coefficients between sensory metrics obtained by Dravnieks and three odor metrics inferred by the LLM as summarized in Table, 
% derived through Mantel's test. 

Table 2 summarizes the correlation coefficients and corresponding p-values from the Mantel tests. 
The p-values obtained in the Mantel tests are influenced by the number of permutations performed; 
in our analysis, the all p-values were 0.001, indicating statistically significant correlations across all metrics.

% The correlation coefficients between sensory metrics and the three inferred odor metrics are summarized in Table 2. 
% The p-values obtained in the Mantel tests are influenced by the number of permutations performed. 
% In our analysis, the minimum p-value was 0.001, 
% suggesting that the correlations presented in Table 2 are statistically significant across all metrics.

While SimByLLM showed a Pearson correlation of 0.332 with DravOrig, GenSE(+EG300m)'s correlation decreased to 0.295. 
This reduction is likely due to variability in writing style within GenSE's LLM output.  
% (e.g., sentences vs. bulleted lists and inclusion of unrelated odor explanations, especially in Gemma3 12B)
% This inconsistency persisted even when using alternative embedding models (see Supplementary Information).
OptGenSE(+EG300m), benefiting from prompt optimization, achieved a correlation coefficient of 0.344 with DravOrig 
- indicating a better level of consistency with the original sensory evaluations. 
The descriptive quality of OptGenSE's outputs in Table 1 also demonstrates that 
prompt refinement leads to more consistent and characteristic odor descriptions.

% Fig. 3 Correlation coefficient of pairwise distance to compare the three inferred representations with the sensory metrics
% \begin{landscape}
\begin{figure} % {figure*}[!t]% 全幅画像挿入
    \centering
    \includegraphics[keepaspectratio, width=3 in ]{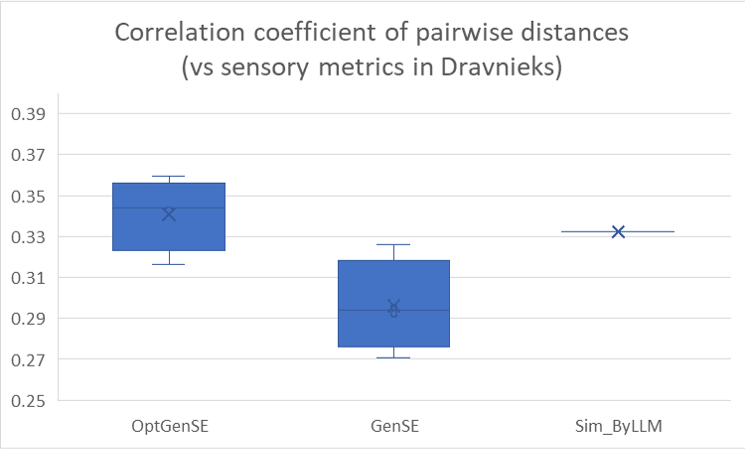}
    \caption{Correlation coefficient of pairwise distance to compare the three inferred representations with Dravnieks' sensory evaluational metrics (DravOrig).
    GenSE and OptGenSE exhibit distinct distributions due to the application of four different embedding models to each set of verbal inferences.}
\end{figure} % {figure*}
% \end{landscape}

To compare the three inferred representations with the sensory metrics, 
we also applied three alternative pre-trained embedding models: 
bge-m3:567m \cite{bge-m3}, mxbai-embed-large:335m \cite{embed2024mxbai}, and nomic-embed-text:latest \cite{nussbaum2024nomic}, 
instead of the embeddinggemma:300m model, then, we performed Mantel tests on their pairwise distances. 
A box plot illustrating the resulting correlation coefficients is shown in Fig. 3. 
In the box plot, the variation across the four embedding models are shown here, whereas no variation of SimByLLM is observed because it does not are any embedding. 
As we switch among the four embedding models, GenSE’s correlation coefficient remains lower than that of SimByLLM, a trend illustrated by the bars.
Notably, OptGenSE achieves an even higher correlation with DravOrig, 
and this result holds regardless of which alternative embedding model is used.

% "C:\Users\yharada\Documents\CurrentTex.260601.DSPyGenSE.acm_vrst2026\fig\CorrCoefBoxP_GPT-4o-mini.png"

\subsection{Stress and the MDS dimension (n-components) } 

% Fig. 4 Stress and the MDS dimension (n-components) for four metrics
% \begin{landscape}
\begin{figure} % {figure*}[!t]% 全幅画像挿入
    \centering
    \includegraphics[keepaspectratio, width=3 in ]{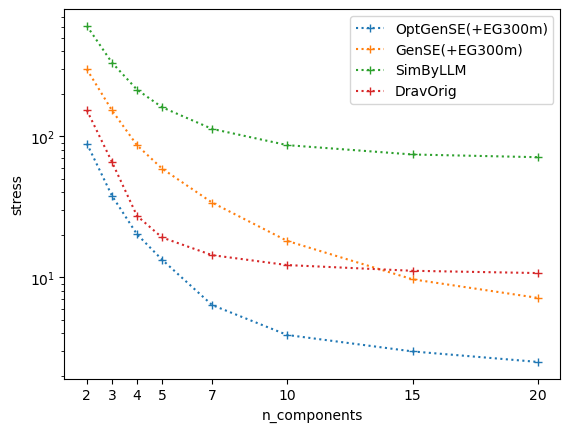}
    \caption{Stress and the MDS dimension (n-components) for four metrics.}
\end{figure} % {figure*}
% \end{landscape}

Fig. 4 illustrates the relationship between stress and MDS dimension. 
We consistently observed that a decreasing trend in stress as the MDS dimension increases 
for three inferred representations (SimByLLM, GenSE, and OptGenSE), even at higher dimensions, 
reaching or even surpassing the levels seen with DravOrig.
This result supports that both OptGenSE and SimByLLM exhibit comparable tendencies in reproducing the sensory metrics of Dravnieks. 

In DravOrig, GenSE and OptGenSE, pairwise distance between the embedding was calculated to make distance matrix for MDS.

% Figure 3 illustrates the relationship between stress and MDS dimension for each odorants. 
% The figure demonstrates a decreasing trend in stress as the MDS dimension increases, even at higher dimensions. 
% Mirroring the observed relationship between reasoning capability and parameter size across four representation. 
% Based on the strong pairwise distance correlation between the Dravnieks data and the Stress-MDS dimension relationship, 
% we conclude that OptGenSE and SimByLLM show comparable abilities in reproducing the sensory metrics present in the Dravnieks dataset.

% Fig. 4 illustrates the relationship between Stress and the MDS dimension for odorants from the Dravnieks dataset. 
% It demonstrates a decrease in stress with MDS dimension, 
% even it the region of high dimension. % when using larger numbers of dimensions. 
% % Our analysis indicates that the reasoning capabilities of LLMs for odor names exhibited a similar trend with model parameter size as observed for odor descriptors. 
% In our analysis, the similar relationship between reasoning capability and parameter size observed with odor descriptors can also be seen with odor names. 
% Based on the substantial pairwise distance correlation between Dravnieks data 
% and the relationship between Stress and the MDS dimension, 
% we concluded that both OptGenSE and SimByLLM exhibit comparable capabilities in reproducing the sensory metrics in Dravnieks. 

\subsection{Odor Maps from LLM-Inference of odor verbal descriptions }

Further optimization may yield results even more consistent with sensory evaluations; 
however, currently, it appears that both OptGenSE and SimByLLM exhibit comparable capabilities in reproducing the original DravSE. 
% , as refering their correlation coefficients. 
Following previous procedures, a multidimensional scaling (MDS) map was constructed from the set of pairwise distance in OptGenSE (Fig. 1).

Figs. 1 and 2 present the odor maps via a 2D representation of MDS based on pairwise similarities from OptGenSE and DravOrig, respectively. 
% MDS can be executed for any dimension size, and the resulting coordinates are readily available for automated fragrance creation
% —a valuable tool for aligning odor space with sensory evaluational coordinates.
Two figures highlight alcohols with light blue, carboxylic acids with yellow, and aromatics with red, respectively. 
In both cases, chemicals with the same functional groups are located closely to each other. 
The two-dimensional MDS mapping of odor names, generated by the two representations, 
shows that odor names within each group possess similar relative locations in both cases.

\section{Conclusion}

In this study, 
we present a method that simultaneously provides verbal descriptions of odors and odor maps. 
We observed that attempting to obtain textual descriptions of odors without prompt optimization resulted in variability in expression patterns, 
leading to inconsistent results with LLM-derived similarity scores. 
However, by optimizing the prompts, we were able to reduce this inconsistency. 
Furthermore, using a Mantel test to compare our results to the sensory metrics reported by Dravnieks, 
we found that the ability by optimized inference was comparable to that of LLM-derived similarity scores.

Large-scale odor map is very essential in the progress of digital scent technology. 
We will extend this work to large-scale map in the near future. 
This odor map holds potential for automated scent creation by providing coordinate transformations of odor space 
in a way consistent with human sensory evaluation.

% \section*{Supporting information}

% % Include only the SI item label in the paragraph heading. Use the \nameref{label} command to cite SI items in the text.
% \paragraph*{Dravnieks odor descriptors.csv.}
% \label{behavior_1.csv.}
% % \textbf{}
% This file provides sensory evaluations, odor descriptors, and ingredient information from the Dravnieks dataset.

% \paragraph*{dravniks stimuli for simbyllm.csv }
% \label{dravniks stimuli for simbyllm.csv }
% % \textbf{ }
% This file contains the ingredients within the Dravnieks dataset, specifically listing odor names that were used as input prompts for the LLMs.

% \paragraph*{essential oils.csv.}
% \label{essential oils.csv}
% % \textbf{}
% This file provides a comprehensive list of 96 essential oils, with a subset of 75 used as input prompts for the LLMs.

\section*{Acknowledgments}

This work was supported by the Japan Science and Technology Agency (JST) Future Society Creation Program (Grant Number JPMJMI25H1).
The funders had no role in study design, data collection and analysis, decision to publish, or preparation of the manuscript.

\bibliographystyle{unsrtnat}
\bibliography{reference}  %%% Uncomment this line and comment out the ``thebibliography'' section below to use the external .bib file (using bibtex) .




% 	\bibitem{keshet2016prediction}
% 	Keshet, Renato, Alina Maor, and George Kour.
% 	\newblock Prediction-Based, Prioritized Market-Share Insight Extraction.
% 	\newblock In {\em Advanced Data Mining and Applications (ADMA), 2016 12th International 
%                       Conference of}, pages 81--94,2016.

% \end{thebibliography}

\end{document}